\documentclass[aps,longbibliography,showpacs,twocolumn,superscriptaddress]{revtex4-1}
\usepackage{amsmath,amssymb,amsfonts,bm}
\usepackage{graphicx}
\usepackage{epstopdf}
\usepackage{dcolumn}
\usepackage{mathrsfs}
\usepackage{bbold}
\usepackage{dsfont}
\usepackage{float}
\usepackage[colorlinks=true,linkcolor=blue,citecolor=blue, urlcolor=blue,bookmarks=false]{hyperref}
\usepackage{tgtermes}
\usepackage{multirow}
\usepackage{bbm}
\usepackage{wasysym}
\usepackage{newtxtext}
\usepackage[varvw]{newtxmath}
\usepackage{hyperref}
\usepackage{array}
\usepackage{bm}
\usepackage{color}
\usepackage{float}
\usepackage{graphicx}
\usepackage{natbib}
\usepackage{newtxtext}
\usepackage{newtxmath}
\usepackage[caption=false]{subfig}
\usepackage{mathrsfs}
\usepackage{amsmath,amssymb}
\usepackage{verbatim}
\usepackage{amsfonts}
\usepackage{dcolumn}
\usepackage{ulem}

\newcommand{\beq}{\begin{eqnarray} }
\newcommand{\eeq}{\end{eqnarray} }
\newcommand{\Beq}{\begin{eqnarray*} }
\newcommand{\Eeq}{\end{eqnarray*} }
\newcommand{\Bmat}{\left(\begin{matrix}}
\newcommand{\Emat}{\end{matrix}\right)}
\newcommand{\up}{\uparrow}
\newcommand{\dn}{\downarrow}

\begin{document}

\title{Effect of ring-exchange interactions in the extended Kitaev honeycomb model}
\author{Jiucai Wang}
\affiliation{Department of Physics and HKU-UCAS Joint Institute for Theoretical and Computational Physics, The University of Hong Kong, Pokfulam Road, Hong Kong, China}
\affiliation{Institute for Advanced Study, Tsinghua University, Beijing 100084, China}

\author{Zheng-Xin Liu}
\email{liuzxphys@ruc.edu.cn}
\affiliation{Department of Physics and Beijing Key Laboratory of Opto-electronic Functional Materials and Micro-nano Devices, Renmin University of China, Beijing, 100872, China}
\affiliation{Key Laboratory of Quantum State Construction and Manipulation (Ministry of Education), Renmin University of China, Beijing, 100872, China}

\date{\today}

\begin{abstract}
Motivated by the possible triple-$\bf Q$ classical order in the Kitaev candidate material Na$_2$Co$_2$TeO$_6$, we 
investigate microscopic models that may stabilize the triple-$\bf Q$ order by studying an extended Kitaev honeycomb model with ring-exchange interactions (namely, the $K$-$\Gamma$-$\Gamma'$-$J_{\rm R}$ model) using the variational Monte Carlo method.    
It turns out that with a positive ring-exchange interaction ($J_{\rm R}>0$) 
there indeed appears an exotic non-coplanar triple-$\bf Q$ ordered state featured by three Bragg peaks at symmetry-related \pmb M points in the crystallographic Brillouin zone.  
A magnetic field in the honeycomb plane can suppress the triple-$\bf Q$ order and induce a gapless quantum spin liquid (QSL) with eight cones.
Furthermore, with the increase of $J_{\rm R}$ a proximate Kitaev spin liquid with 8 Majorana cones labeled “PKSL8” is found, which is very stable over a large range of $\Gamma$ interactions. The PKSL8 state shares the same projective symmetry group with the Kitaev spin liquid (KSL) which is located at small $\Gamma$ and $J_{\rm R}$. In a weak magnetic field applied normal to the honeycomb plane, the PKSL8 turns into an Abelian chiral spin liquid with Chern number $\nu$$=$$-4$, unlike the KSL which yields a chiral spin liquid with $\nu$$=$$1$. Since the triple-$\bf Q$ phase is adjacent to two QSLs in the phase diagram, our work suggests that it is more hopeful to experimentally realize the exotic QSL phases starting from the triple-$\bf Q$ order.
\end{abstract}

\maketitle

\section{Introduction}
Quantum spin liquids (QSLs) are exotic phases of matter exhibiting no conventional long-range order even at zero temperature due to strong quantum fluctuations\cite{Anderson1973,Balents,Balents2016,zhouyi,Knolle2019,Senthil2020}.
QSLs are characterized by long-range entanglement and the existence of intrinsic fractional excitations\cite{XGWen,Levin2006,Kitaev2006,Chen2010}.
These fractionalized quasiparticles can obey fractional 
exchanging statistics and are dubbed anyons. 
However, it is challenging to construct lattice spin models in two or higher dimensions to support spin-liquid ground states.
After Kitaev proposed a honeycomb-lattice model of bond-dependent Ising-type spin interactions ($S^\gamma_i S_j^\gamma$) supporting an exactly solvable spin-liquid ground state\cite{Kitaev}, much progress has been made in systems with strong spin-orbit coupling\cite{Xiang2007,Yao2007,Yao2009,Yao2011,Tu2020,Zhou2020,Ma2022}.
The spin-liquid phase of the Kitaev model has a gapless or gapped excitation spectrum, and in a suitable magnetic field, the gapless Kitaev spin liquid (KSL) can be turned into a gapped chiral spin liquid (CSL) that supports non-Abelian anyonic excitations.

To realize the KSL, a series of spin-orbit entangled candidate materials have been proposed and studied profoundly\cite{rjk,rcjk}, such as $\alpha$-RuCl$_3$\cite{rfgft,YJKim,rsea,rjea,rcaoetal}, $\alpha$-Li$_2$IrO$_3$\cite{incomm,Coldea2019}, Na$_2$IrO$_3$\cite{singh,XLiu,ryea,rchoietal}, Cu$_2$IrO$_3$\cite{abra,Choi2019}, H$_3$LiIr$_2$O$_6$\cite{rhliiro} and Na$_2$Co$_2$TeO$_6$\cite{lefran,bera,LiYuan2020,YaoW2022,Stock2020,Ma2021,Garlea2021,Rachel2022}.
However, most of these materials manifest a magnetic long-range order at low temperatures, instead of having a spin-liquid ground state, indicating the existence of other symmetry-allowed interactions beyond the Kitaev coupling such as the nearest-neighbor symmetric off-diagonal $\Gamma$ interaction ($S^\alpha_iS_j^\beta+S^\beta_iS_j^\alpha$) and Heisenberg interaction ($\pmb S_i\cdot\pmb S_j$).
To explain the experimental data in Kitaev materials, a series of lattice spin models have been proposed. 
For instance, a third-neighbor Heisenberg interaction\cite{Vojta2017,Winter2017,Kimchi2011,Winter2016} or a nearest-neighbor off-diagonal $\Gamma'$ interaction ($S^\alpha_iS_j^\gamma + S^\gamma_iS_j^\alpha + S^\beta_iS_j^\gamma + S^\gamma_i S_j^\beta$)\cite{Kee2014,Kee2019,Kim2020} is included in the effective spin model to interpret the zigzag order at low temperatures.

However, the experimental results of inelastic neutron scattering in Na$_2$Co$_2$TeO$_6$
seem to indicate a triple-$\bf Q$ ordered state\cite{LiYuan2021,LiYuan2022}, which supports three Bragg peaks at symmetry-related points in the Brillouin zone.
Thus, it is difficult to distinguish between the triple-$\bf Q$ order and the zigzag order in neutron diffraction experiments due to domain mixing effects.
The triple-$\bf Q$ order has three-fold rotation symmetry $C_3^*$ (both in spin and lattice space), while the zigzag order breaks this symmetry.
In some sense, the $C_3^*$ symmetric triple-$\bf Q$ order can be obtained by superposing three zigzag order parameters.
An important question is how to stabilize this triple-$\bf Q$ order in a lattice model.
A simple idea is to add some $C_3^*$ symmetric exchange interactions in the proposed lattice model.
For instance, in a magnetic field applied normal to the honeycomb plane, the Heisenberg-Kitaev honeycomb model may support a non-coplanar triple-$\bf Q$ state\cite{Vojta2016,Andrade2020}.
In addition, near the hidden-SU(2) symmetric point in an extended $K$-$\Gamma$-$\Gamma'$-$J$ model, the negative nonbilinear six-spin exchange interaction favors triple-$\bf Q$ order\cite{Lukas2022}.
Here we propose another mechanism to stabilize this triple-$\bf Q$ order in the generic extended Kitaev honeycomb model (which does not need hidden-SU(2) symmetry): multi-electron ring-exchange interactions ($J_{\rm R}$).
Generally, the ring-exchange interaction possibly arises from higher-order corrections in the strong-coupling expansion of the Hubbard model\cite{MacDonald1988,Yang2012}.
Additionally, a triple-$\bf Q$ order is also stabilized by higher-order interactions in higher spin systems\cite{Motome2023}.

\begin{figure*}[t]
\begin{flushleft}
\includegraphics[width=17.8cm]{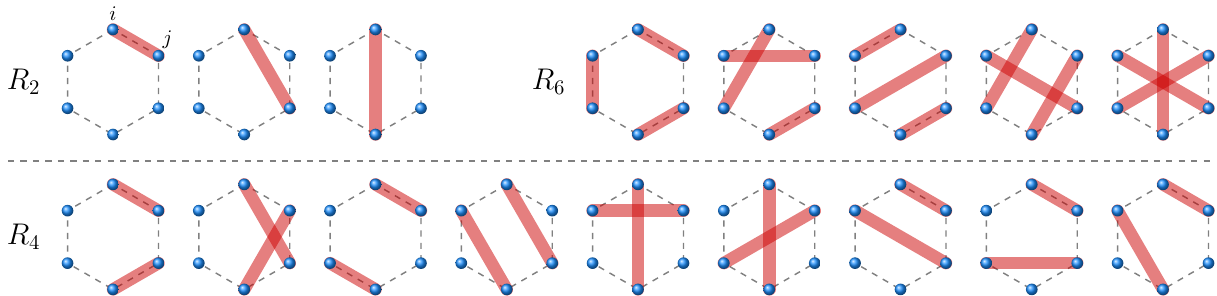}
\end{flushleft}
\caption{Schematics of one-pair ($R_2$, left upper row), three-pair ($R_6$, right upper row), and two-pair ($R_4$, bottom row) spins exchange interactions on each plaquette of the honeycomb lattice, respectively. In the one-pair case, the number of elements is 6, 6, and 3, respectively. In the two-pair case, the number of elements is 6, 6, 3, 3, 6, 3, 6, 6, and 6, respectively. In the three-pair case, the number of elements is 2, 6, 3, 3, and 1, respectively. The light red shadow connecting one pair of sites ($i,j$) represents the Heisenberg interaction ($\pmb S_i\cdot\pmb S_j$).}
\label{Ring}
\end{figure*}

In this paper, we investigate the quantum $K$-$\Gamma$-$\Gamma'$-$J_{\rm R}$ model using the variational Monte Carlo (VMC) approach, and the global phase diagram is obtained.
The positive ring-exchange interaction ($J_{\rm R}>0$) indeed supports a triple-$\bf Q$ ordered state that features Bragg peaks at three $\rm \bf M$ points in the crystallographic Brillouin zone.
In addition, the triple-$\bf Q$ order is stable under weak magnetic fields. Interestingly, the magnetic field along $(\bf {x-y})$ direction may suppress the triple-$\bf Q$ order and induce a gapless Z$_2$ QSL with 8 cones on the high symmetry line of the first Brillouin zone.
This result is instructive for an experimental search of gapless QSLs and triple-$\bf Q$ order in related materials.
Furthermore, with the increase of the ring-exchange interaction with $J_{\rm R}/|K| \gtrsim 0.15$, a spin-liquid phase is found which is much stabler than the KSL phase.
It contains 8 Majorana cones in its spinon excitation spectrum and shares the same projective symmetry group (PSG)\cite{igg,You_PSG} with the KSL, and is thus called the proximate-KSL8 (PKSL8) phase. 
In a magnetic field applied normal to the honeycomb plane, the PKSL8 realizes an Abelian CSL with Chern number $\nu$$=$$-4$ while the KSL is turned into a non-Abelian CSL with $\nu$$=$$1$.

The rest of the paper is organized as follows: In Sec.\ref{Model}, we introduce the $K$-$\Gamma$-$\Gamma'$-$J_{\rm R}$ model on the honeycomb lattice and describe the numerical method used in this work. In Sec.\ref{PhaseDiagram}, we summarize our main findings in the phase diagram including various QSLs and magnetically ordered phases.
We further show the physical response of the system to magnetic fields for QSL
phases as well as magnetically ordered phases in Sec.\ref{Field}.
The paper is concluded in Sec.\ref{Conclusion}.

\begin{figure}[b]
\includegraphics[width=8.5cm]{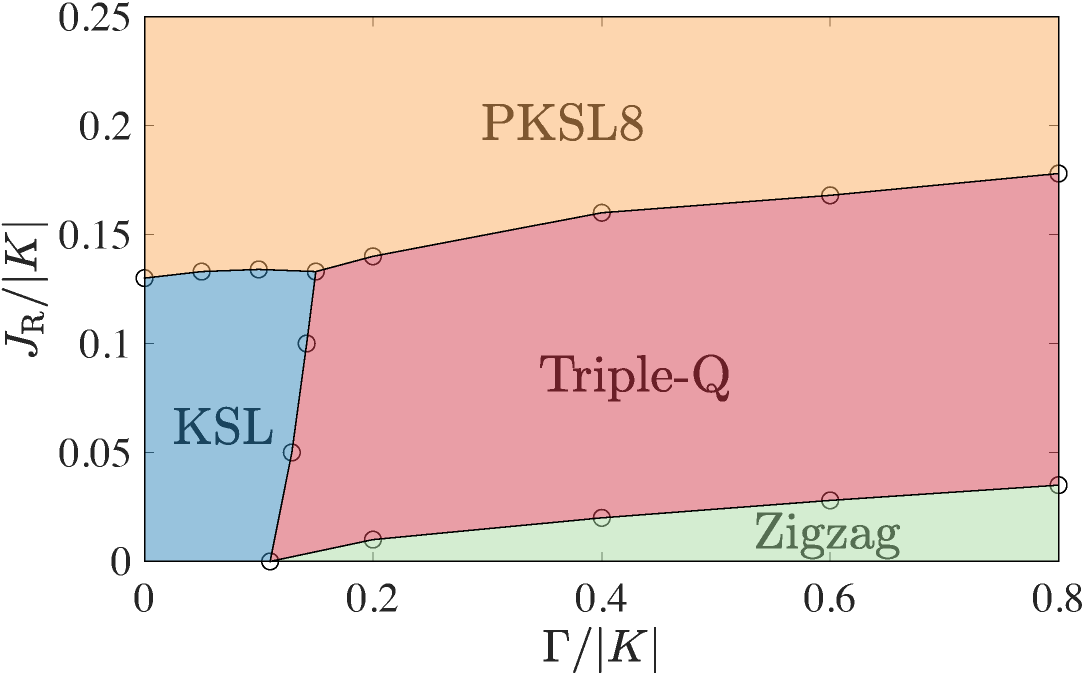}
\caption{ Phase diagram of the quantum $K$-$\Gamma$-$\Gamma'$-$J_{\rm R}$ model for $K < 0$, $\Gamma >0$, $J_{\rm R}>0$, and $\Gamma'/|K|=-0.05$ in the limit of large system size. 
There are two QSL phases of different types but with the same PSG, the KSL and the PKSL8. 
The magnetically ordered phases include zigzag and triple-$\bf Q$ order.}
\label{KGammaGammaPrimeRing}
\end{figure}

\begin{figure*}[t]
\includegraphics[width=7.0cm]{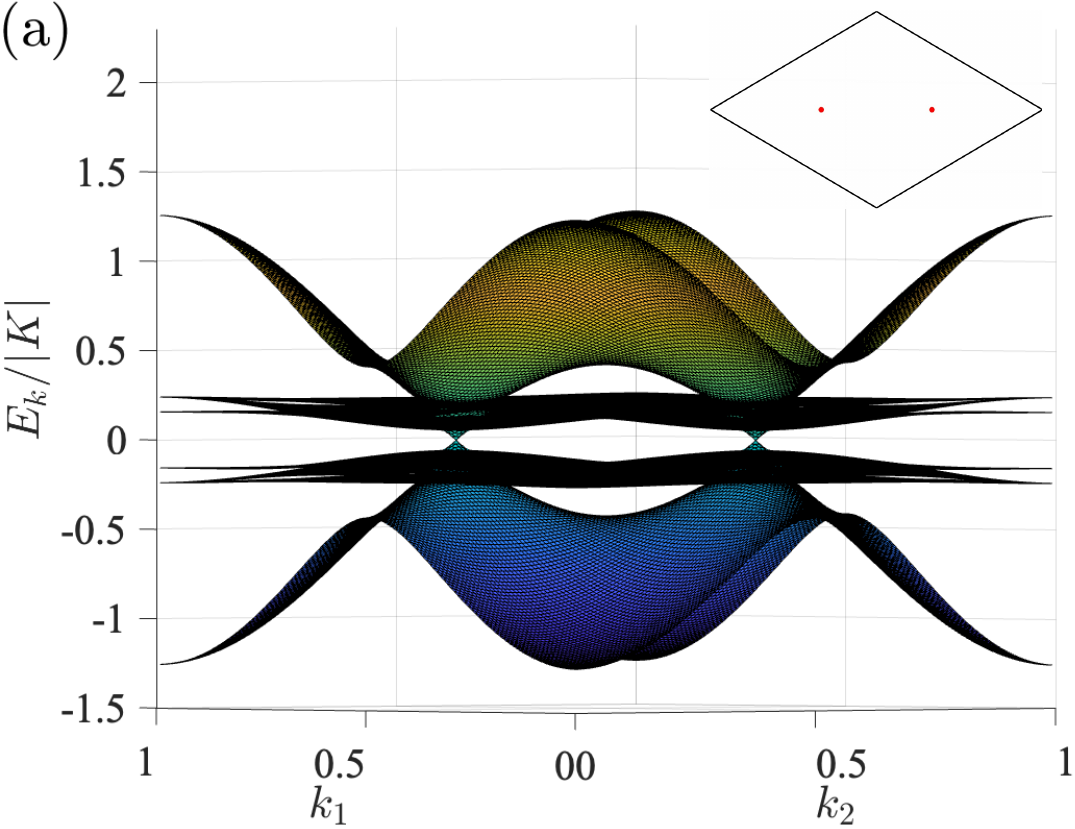}
\includegraphics[width=7.0cm]{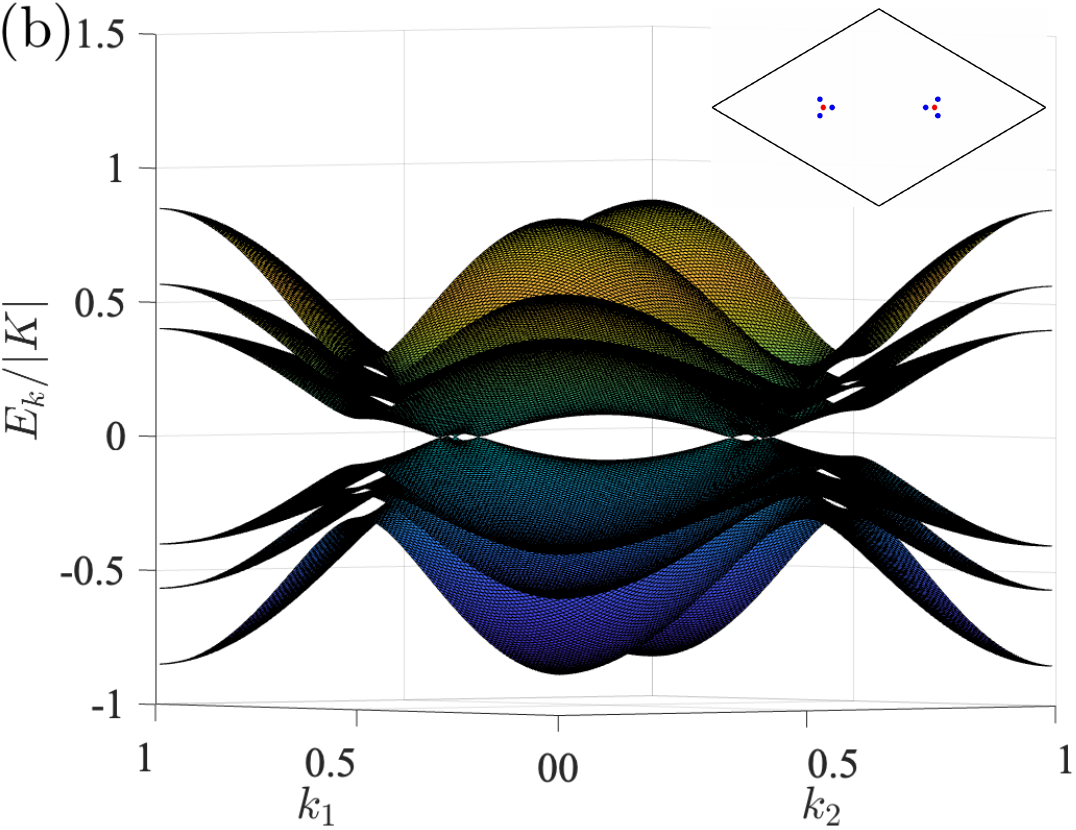}
\includegraphics[width=3.4cm]{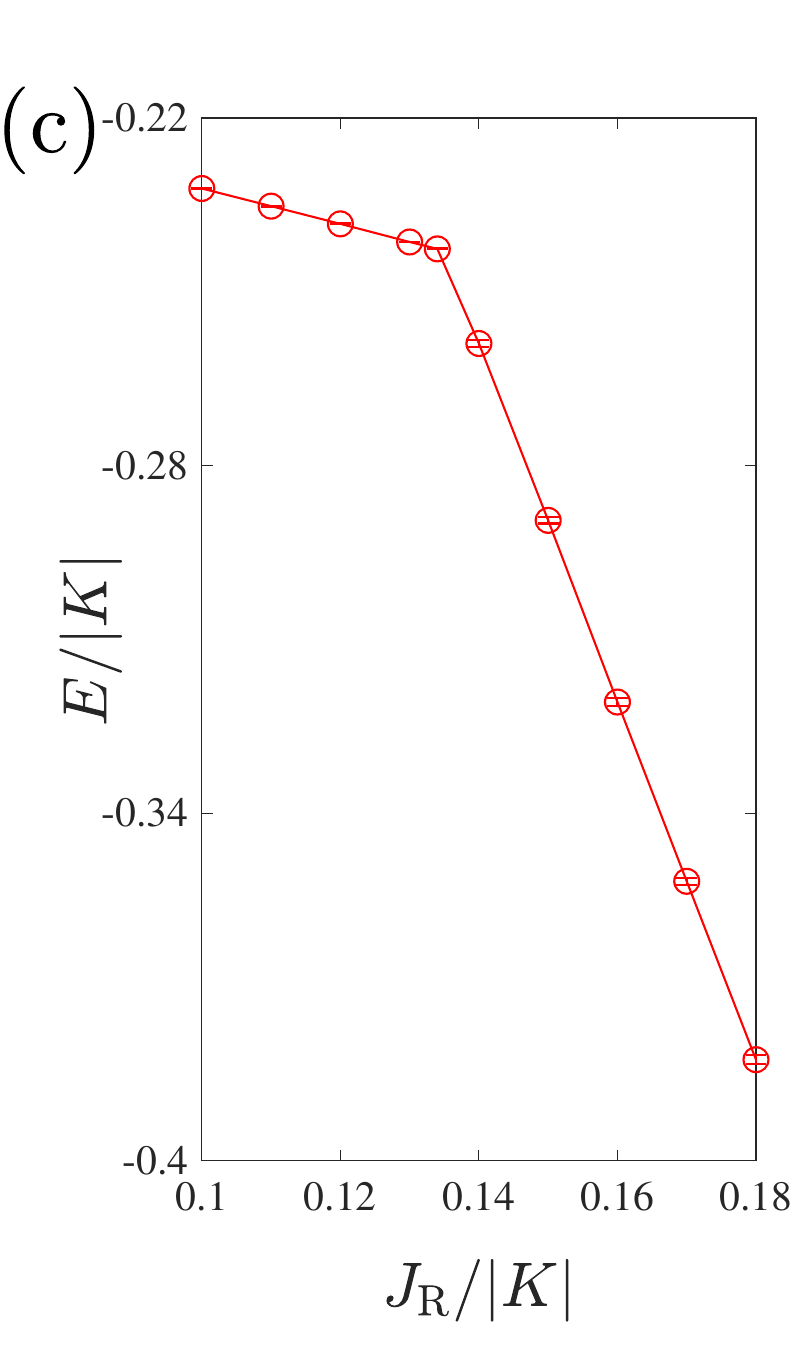}
\caption{ (a) Spinon dispersion in the KSL, drawn with $\Gamma/|K| = 0.1$, $\Gamma'/|K|=-0.05$, $J_{\rm R}/|K|=0.1$, showing 2 Majorana cones. 
(b) Spinon dispersion in the PKSL8, drawn with $\Gamma/|K| = 0.1$, $\Gamma'/|K|=-0.05$, $J_{\rm R}/|K|=0.15$, showing 8 Majorana cones.
(c) Ground-state energy per site of the $K$-$\Gamma$-$\Gamma'$-$J_{\rm R}$ model at fixed $\Gamma/|K| = 0.1$, $\Gamma'/|K|=-0.05$, showing a clear first-order phase transition.}
\label{Dispersions}
\end{figure*}

\section{Model and method}\label{Model}
We start with the extended Kitaev model on the honeycomb lattice containing the Kitaev-exchange ($K$-term), off-diagonal exchange ($\Gamma$-, and $\Gamma'$-term), and ring exchange ($J_{\rm R}$-term) interactions:
\begin{eqnarray}\label{KGGR}
&&H = H_K+ H_{\Gamma} + H_{\Gamma'} + H_{\rm R}, 
\end{eqnarray}
with
\begin{eqnarray}
&&H_K=\sum_{\langle i,j \rangle \in\alpha\beta(\gamma)}  K S_i^\gamma S_j^\gamma,\nonumber \\
&&H_{\Gamma}= \sum_{\langle i,j \rangle \in\alpha\beta(\gamma)} \Gamma (S_i^\alpha S_j^\beta + S_i^\beta S_j^\alpha), \nonumber \\
&&H_{\Gamma'}= \sum_{\langle i,j \rangle \in\alpha\beta(\gamma)} \Gamma' (S_i^\alpha S_j^\gamma + S_i^\gamma S_j^\alpha + S_i^\beta S_j^\gamma + S_i^\gamma S_j^\beta), \nonumber \\
&&H_{\rm R} = \sum_{\hexagon} J_{\rm R} \hat P_{\hexagon} , \nonumber 
\end{eqnarray}
where $\langle i,j \rangle$ denotes nearest-neighbor sites, $\gamma$ labels the type of the bond $\langle i,j \rangle$ on the honeycomb lattice, $\alpha$, $\beta$, and $\gamma$ stand for the spin indeces; and $\hat P_{\hexagon}=-{3\over 8} + R_2+R_4+R_6$ stands for the ring-exchange interaction on the hexagon whose explicit form is given in Eqs.(\ref{P6})\&(\ref{app:ring}). Additionally, its graphical form in the spin basis is represented in Fig.\ref{Ring}.

In most Kitaev materials, the Kitaev coupling has a negative sign ($K < 0$). In the present work, we adopt the parameters of the interactions such that $K<0$, $\Gamma>0$, $\Gamma'<0$, and $J_{\rm R} >0$. Due to spin-orbit coupling, the symmetry of the model is described by the finite magnetic point group $D_{3d} \times Z_2^T$ besides lattice translation symmetries, where $Z_2^T = \{E,T\}$ is the time reversal group.

The model (\ref{KGGR}) can be mapped into an interacting fermionic model in the Majorana representation $S_i^m=ib_i^mc_i$ (under the constraint $b_i^xb_i^yb_i^zc_i=1$) introduced by Kitaev. One can combine the Majorana fermions into complex fermionic spinons $C_i=(c_{i \up}, c_{i \dn})^T$ such that the constraint is mapped into the particle-number constraint $C_i^\dag C_i=1$. In the complex fermion representation, the ring-exchange interaction can be simply expressed as
\begin{eqnarray}\label{P6}
\hat P_{\hexagon}=   -\hat\chi_{ij}\hat\chi_{jk}\hat\chi_{kl}\hat\chi_{lm}\hat\chi_{mn}\hat\chi_{ni} - {\rm cyclic}(ijklmn) +  {\rm H.c.}, 
\end{eqnarray}
where the index $i,j,k,l,m,n \in {\rm hexagon}$ (sorting clockwise on the hexagon), and $\hat\chi_{ij}=C_i^\dagger C_j=c_{i\uparrow}^\dagger c_{j\uparrow} + c_{i\downarrow}^\dagger c_{j\downarrow}$.

The ground-state energy of the model can be calculated from VMC using the Gutzwiller projected mean-field ground states as trial wave functions. 
Thus we perform Gutzwiller projection to the mean-field ground state $|\Psi_{\rm mf} (\pmb R)\rangle$ to ensure the particle number constraint. The projected states $|\Psi (\pmb R)\rangle = P_G |\Psi_{\rm mf}(\pmb R) \rangle$ provide a series of trial wave functions depending on the choice of the mean-field Hamiltonian $H_{\rm mf}(\pmb R)$, where $P_G$ denotes a Gutzwiller projection and $\pmb R$ are treated as variational parameters.  The energy of the trial state $E (\pmb R) = \langle \Psi(\pmb R) |H| \Psi(\pmb R) \rangle / \langle \Psi(\pmb R)| \Psi(\pmb R) \rangle$ is computed using Monte Carlo sampling, and the optimal parameters $\pmb R$ are determined by minimizing the energy $E(\pmb R)$.

In constructing the mean-field Hamiltonian, we follow the guidance of the PSG and construct different types of QSL ansatz as trial states (for details see Appendix \ref{Gutzwiller}). The magnetic order is treated as a background field, in which the ordering pattern is obtained from single-$\bf Q$ or multi-$\bf Q$ approximation, and the amplitude is determined by minimizing the energy. 

Our VMC calculations are performed on a lattice with 8 by 8 unit cells, namely, 128 sites.
The spinon dispersion of the QSLs can be qualitatively obtained by diagonalizing the mean-field Hamiltonian in a larger system size with the optimized parameters from VMC calculations. From this dispersion, we can locate the positions of the nodes in the gapless QSLs.

\section{Phase diagram}\label{PhaseDiagram}
Because theoretical studies have shown that a very small negative $\Gamma'$ interaction can support a zigzag-ordered ground state\cite{Kee2019,Kim2020,Wang2020}, we study the $K$-$\Gamma$-$\Gamma'$-$J_{\rm R}$ model at fixed $\Gamma'/|K|=-0.05$ for simplicity using the VMC approach.
Figure \ref{KGammaGammaPrimeRing} shows the VMC phase diagram of this quantum model.
We find that two spin-liquid states are robust from our VMC calculations.
One is the KSL, whose regime 
is bounded approximately by $\Gamma/|K| = 0.15$ and $J_{\rm R}/|K| = 0.13$.
The second one is PKSL8 which is one QSL proximate to the KSL, whose regime of stability is $J_{\rm R}/|K| \gtrsim 0.15$.
The KSL and PKSL8 have the same PSG despite being physically quite different states. 
In contrast to the spinon excitation spectrum of the KSL, which has two
Majorana cones in the first Brillouin zone [Fig.\ref{Dispersions}(a)], the PKSL has eight Majorana cones [Fig.\ref{Dispersions}(b)].
These cones are protected from local perturbations by the combination of spatial inversion and time reversal symmetry.
Our numerical calculation strongly suggests the phase transition between KSL and PKSL8 is sharply first-order due to ground-state energy level crossing, as shown in Fig.\ref{Dispersions}(c).

\begin{figure}[t]
\begin{flushleft}
\includegraphics[width=2.97cm,height=1.8cm]{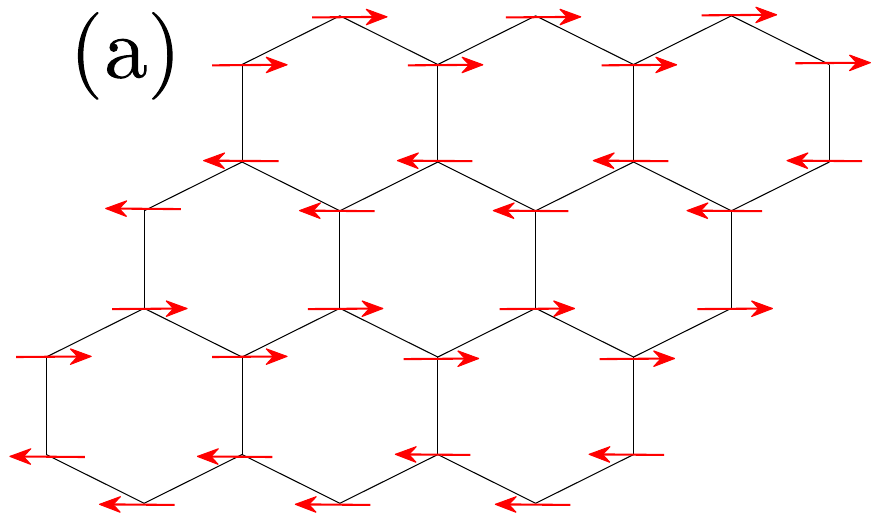}
\includegraphics[width=2.76cm,height=1.8cm]{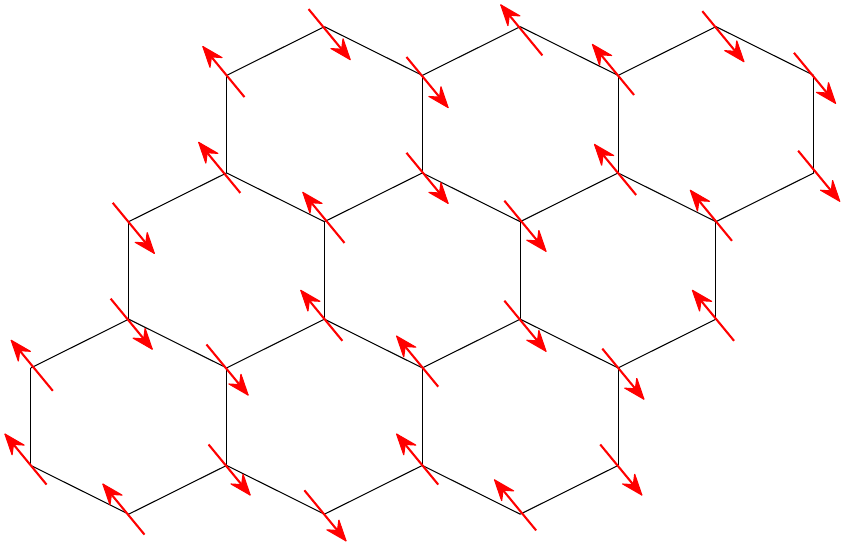}
\includegraphics[width=2.76cm,height=1.8cm]{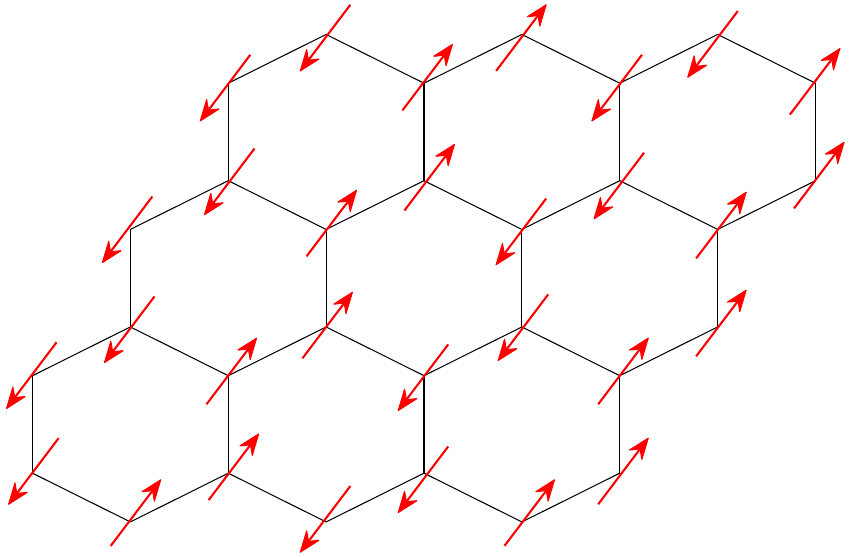}\\  
\end{flushleft}
\includegraphics[width=4.0cm,height=3.2cm]{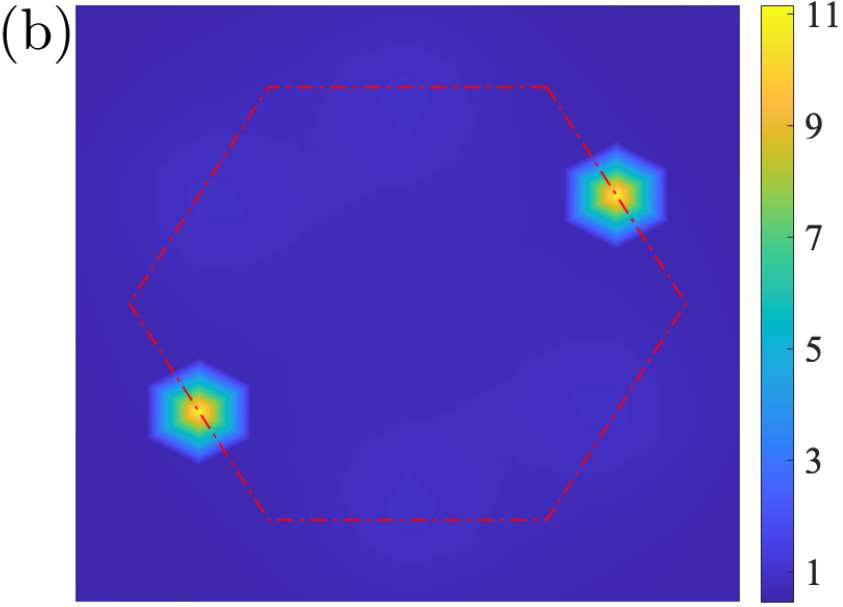}
\includegraphics[width=4.0cm,height=3.2cm]{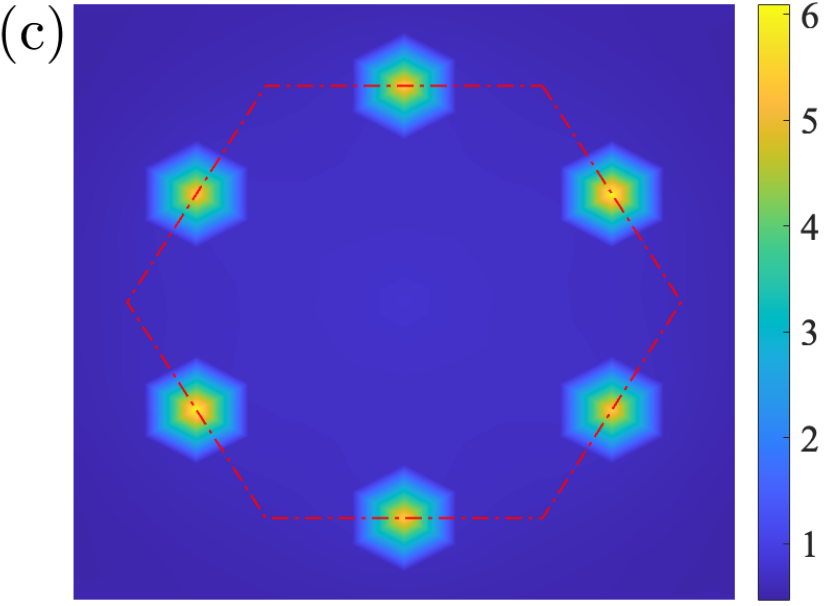}
\caption{(a) Schematics of three-type zigzag orders connected by a threefold rotation $C_3^*$ (both in spin and lattice space). The triple-$\bf Q$ order is formed by the vector sum of three $C_3^*$-related zigzag structures.
(b) The static spin structure factor $S(\pmb{q})$ for zigzag order (with $\Gamma/|K|=0.4$, $\Gamma'/|K|=-0.05$).
(c) The static spin structure factor $S(\pmb{q})$ for triple-$\bf Q$ order (with $\Gamma/|K|=0.4$, $\Gamma'/|K|=-0.05$, $J_{\rm R}/|K|=0.1$).
The red dotted line represents the first crystallographic Brillouin zone, and bright points in the figures are $\rm \bf M$ points.}
\label{ZZPattern}
\end{figure}

Besides the QSL phases, there are two magnetically ordered states in a large region of the phase diagram (for details see Fig.\ref{KGammaGammaPrimeRing}).
One is a zigzag-ordered state, and the other is a triple-$\bf Q$ ordered state.
The spin configuration of the zigzag order is shown in Fig.\ref{ZZPattern}(a).
The triple-$\bf Q$ order with an 8-site magnetic unit cell is formed by superposing three zigzag order parameters\cite{LiYuan2021}.
Therefore, the triple-$\bf Q$ order has three-fold rotation symmetry $C_3^*$ (including both spin rotation and lattice rotation).
We find that the positive ring-exchange interaction may restore $C_3^*$ symmetry due to quantum fluctuations, resulting in the emergence of the triple-$\bf Q$ state.
To check the magnetic properties of the projected state, we compute the static structure factor,
\begin{eqnarray}
S(\pmb{q})= \frac{1}{N}\sum_{i,j}e^{i\pmb{q} \cdot (\pmb{r}_i-\pmb{r}_j)}\langle \pmb S_{\pmb{r}_i}\cdot \pmb S_{\pmb{r}_j}\rangle.
\end{eqnarray}
As shown in Fig.\ref{ZZPattern}, for the triple-$\bf Q$ state, there are three inequivalent Bragg peaks in the first crystallographic Brillouin zone, whereas there is one inequivalent Bragg peak in the zigzag order.
However, for spin-liquid states there are no prominent peaks in the static structure factor (not shown).

In principle, the ordered phase and spin-liquid phase could be separated by a second-order phase transition or even by an intermediate phase with coexisting magnetic and Z$_2$ topological order. However, according to our calculations, the phase transitions are of first order, as can be observed either in the level crossing in the ground-state energy or through the discontinuities in the optimal variational parameters.

\begin{figure}[b]
\includegraphics[width=8.5cm]{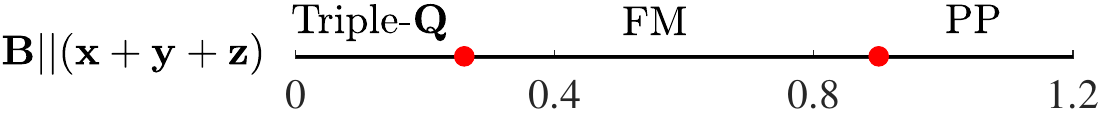}\\  \hspace*{\fill} \\
\includegraphics[width=8.5cm]{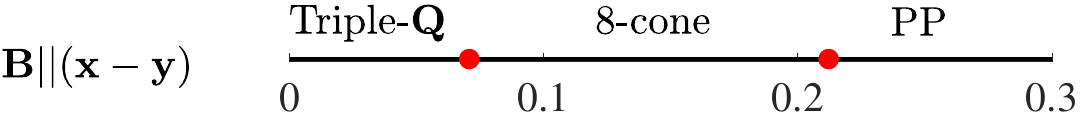}\\  \hspace*{\fill} \\
\includegraphics[width=8.5cm]{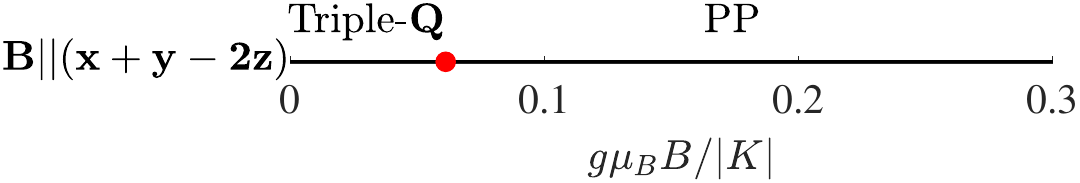}
\caption{Phase diagrams of the triple-$\bf Q$ state ($\Gamma/|K| = 0.4$, $\Gamma'/|K| = -0.05$, $J_{\rm R}/|K|=0.1$) in a magnetic field applied in the $(\bf {x+y+z})$ direction, in the $(\bf {x-y})$ direction, and in the $(\bf {x+y-2z})$ direction. “8-cone” denotes a phase whose low-energy spinon dispersion has eight cones on the high symmetry line (i.e., the horizontal line in the first Brillouin zone).}
\label{TriplQField}
\end{figure}

An interesting observation is that the triple-$\bf Q$ phase is adjacent to two QSL phases at $J_{\rm R}> 0$. 
This indicates that the triple-$\bf Q$ phase is `closer' to QSLs than the zigzag phase, and thus provides a helpful hint for experimental searching of QSLs. Especially, starting from the triple-$\bf Q$ phase, one can hopefully realize the exotic PKSL8 phase (by tuning the interactions in certain materials) which is neighboring to the triple-$\bf Q$ phase in a large parameter region.

\section{Effect of magnetic fields}\label{Field}
One of the most exciting properties of the exactly solvable KSL ground state is that it becomes a gapped non-Abelian CSL in a generic magnetic field\cite{Kitaev}. Thus the KSL in a field becomes a $\nu$$=$$1$ CSL, where the non-Abelian statistics arise due to unpaired Majorana modes associated with the vortices.
Our VMC results verify that in a small magnetic field along the $(\pmb x+\pmb y+\pmb z)$ direction these modes are also present in the KSL phase (deviating from the exactly solvable point).

In the PKSL8, there are four pairs of cones in the spinon dispersion, and each pair is connected by inversion or time reversal symmetry.
There is one pair at $\bf K$ and $\bf K'$ points, and three cones near the $\bf K$-point are connected by threefold rotation.
Each of the four pairs of cones becomes gapped in a magnetic field $\pmb B \parallel (\pmb x+\pmb y+\pmb z)$ and contributes a Chern number $\nu$$=$$\pm 1$.
From our VMC calculations, the PKSL8 becomes a $\nu$$=$$-4$ Abelain CSL in a small field, whose nontrivial topological excitations include $e$, $m$ and $\epsilon$.
$e$ and $m$ are the two different types of vortex in the Abelian CSL, which are both semions.
And the $\epsilon$ is the fermion.
Whether there are other spin liquids at a higher magnetic field and the field along other directions is left for future study.
In a CSL with Chern number $\nu$, there are $\nu$ branches of chiral Majorana edge states, each of which
contributes to a total chiral central charge $c_{-} = \nu /2$.
We can obtain the quantized number $\nu$ from the thermal Hall conductance.
The above result is verified within our VMC analysis by calculating the ground state degeneracy (GSD) on a torus, which matches the number of topologically distinct quasiparticle types (for details see Appendix \ref{GSD}).

Now we focus on the response of the ordered phase to magnetic fields.
We only consider the triple-$\bf Q$ state since the region of the zigzag ordered phase is very small.
In the field along the $(\pmb x+\pmb y+\pmb z)$ direction, the triple-$\bf Q$ order is suppressed by the field after which the system enters the ferromagnetic (FM) phase, as shown in Fig.\ref{TriplQField}. 
Then the system enters the polarized phase (PP) through a first-order phase transition in a larger field.
The main difference between FM and PP is that the direction of the spins in the FM phase obviously deviates from the direction of the field (it lies roughly on the plane spanned by $[11\bar{2}]$ and $[111]$ with small canting), while in the polarized phase the spins are aligned along the field direction. From the symmetry point of view, the FM state spontaneously breaks the rotational reflection symmetry ($S_6=C_6M_{111}$), while the polarized phase preserves all the symmetry of the system and contains no symmetry-breaking order.
Since we have only considered a limited number of ansatz in our VMC approach, other methods may be needed to study this phase diagram.
Interestingly, in the field with $\pmb B \parallel (\pmb x-\pmb y)$, there is a direct phase transition from the triple-$\bf Q$ order to the gapless Z$_2$ QSL with 8 Majorana cones on the high symmetry line of the first Brillouin zone.
Note that the critical field in the $(\pmb x-\pmb y)$ direction is smaller than that in the $(\pmb x+\pmb y+\pmb z)$ direction.
To understand the nature of the field-induced gapless QSL, we restore the $D_{3d} \times Z_2^T$ symmetry by removing the magnetic field manually while keeping all the other variational parameters intact, and we find that the field-induced 8-cone state becomes a 20-cone proximate-KSL (PKSL20) state\cite{Wang2020,PKSL}.
However, in the field with $\pmb B \parallel (\pmb x+\pmb y-2\pmb z)$, there is a direct phase transition from the triple-$\bf Q$ order to the polarized phase.
Therefore, we find that different directional fields in the honeycomb plane can have completely different physical consequences, and that the field along the bond direction may induce a gapless spin-liquid phase with Majorana cones.

\section{Conclusion and Discussion}\label{Conclusion}
\begin{figure}[t]
\includegraphics[width=8.6cm]{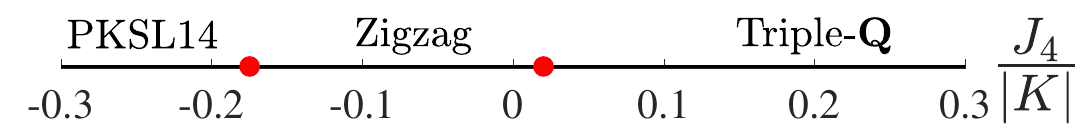}
\caption{Phase diagram of the quantum $K$-$\Gamma$-$\Gamma'$-$J_{\rm 4}$ model with fixing $\Gamma/|K|=0.4$, $\Gamma'/|K|=-0.05$. 
The zigzag ordered state is sandwiched by triple-$\bf Q$ order and PKSL14 state. 
“PKSL14” denotes a proximate KSL state whose low-energy spinon dispersion has 14 Majorana cones in the first Brillouin zone.}
\label{4spinPD}
\end{figure}

Although in a magnetic field applied normal to the honeycomb plane the Heisenberg-Kitaev model supports a non-coplanar triple-$\bf Q$ state, we find the field does not favor the triple-$\bf Q$ state in the $K$-$\Gamma$-$\Gamma'$ model.
Potentially the ring exchange is important to stabilize the triple-$\bf Q$ order in a generic extended Kitaev honeycomb model.
Because the ring-exchange interaction ($\hat P_{\hexagon}$) has SU(2) rotation symmetry and lattice symmetry, we can represent it in the spin basis, for example, one-pair [($\pmb S_i \cdot \pmb S_j$)], two-pair [($\pmb S_i \cdot \pmb S_j$)($\pmb S_k \cdot \pmb S_l$)], and three-pair [($\pmb S_i \cdot \pmb S_j$)($\pmb S_k \cdot \pmb S_l$)($\pmb S_m \cdot \pmb S_n$)] spins exchange interactions on the hexagon, which are represented by diagrams in Fig.\ref{Ring}.
Thus the ring-exchange interaction contains 17 types of spin interactions, and the specific coefficients are given by Eq.(\ref{app:ring}). 
Interestingly, there is a sign structure in two-pair spins exchange ($R_4$) and three-pair spins exchange ($R_6$), i.e., +1(-1) for an even (odd) number of transpositions between light red shadows in Fig.\ref{Ring}.
We guess that the triple-$\bf Q$ order can be stabilized by some of the exchange interactions mentioned above (two special cases are discussed in Appendix \ref{R4}). 
In the following, we mainly consider the nonbilinear two-pair spins exchange interaction,
\begin{eqnarray}\label{4spinHami}
&&H_\text{4-spin} = J_4 \sum_{i,j,k,l,m,n \in \hexagon } R_4(ijklmn),
\end{eqnarray}
where the explicit form of $R_4(ijklmn)$ can be found in Appendix \ref{app:P6}.
From our VMC calculations of the quantum $K$-$\Gamma$-$\Gamma'$-$J_{\rm 4}$ model (i.e., $H'=H_{\rm K}+ H_{\Gamma} + H_{\Gamma'} + H_\text{4-spin}$), we find that the two-pair spins exchange interaction certainly stabilizes the triple-$\bf Q$ order whose regime of stability is $J_{\rm 4}/|K| \gtrsim 0.02 $ at $\Gamma/|K|=0.4$, as shown in Fig.\ref{4spinPD}.
Interestingly, the two-pair spins exchange interaction $J_{\rm 4}/|K| \lesssim -0.175$ may induce a proximate KSL phase with 14 cones (namely, PKSL14) which is similar to our previous work\cite{PKSL}.
For instance, in a weak magnetic field applied normal to the honeycomb plane, the PKSL14 phase turns into a non-Abelian CSL with Chern number $\nu$$=$$5$.

In conclusion, we have studied the quantum $K$-$\Gamma$-$\Gamma'$-$J_{\rm R}$ model on the honeycomb lattice using the variational Monte Carlo method.
One proximate Kitaev spin liquid state with 8 Majorana cones (PKSL8) is found, which shares the same projective symmetry group with the Kitaev spin liquid.
In a suitable field, it turns into an Abelian chiral spin liquid with Chern number $\nu$$=$$-4$.
Interestingly, the positive ring-exchange interaction also induced an exotic triple-$\bf Q$ ordered state which is stable in small magnetic fields. Finally, we propose a mechanism to stabilize this triple-$\bf Q$ order in the generic extended Kitaev honeycomb model.

In addition, there is another type of triple-$\bf Q$ order which can be obtained by superposing three $C_3^*$ symmetry-related stripe orders. Classical simulations\cite{LiYuan} indicate that this stripe-type triple-$\bf Q$ order can be stabilized by tuning the interaction parameters (for instance by adding an antiferromagnetic Heisenberg term \cite{singleQ}). We leave this for future study.

\section{Acknowledgments}
We are grateful to Yuan Li, Wenjie Chen, Lukas Janssen, Qi-Rong Zhao, and Chenjie Wang for valuable discussions,
and we especially thank Yuan Li for insightful suggestions and comments on the manuscript.
J.W. thanks Chun-Jiong Huang for helping to improve the quality of the figures.
This work was supported by the NSF of China (Grants No. 11974421 and No. 12134020), National Key Research and Development Program of China (Grant No. 2022YFA1405301), and Research Grants Council of Hong Kong (Grant No. GRF 11300819).

\appendix

\section{Explicit form of the ring-exchange interaction}\label{app:P6}
Although the ring-exchange interaction can be simply expressed in the complex fermion representation, its form is very complicated in the spin basis.
In the following, we will give the specific form of the ring-exchange interaction,
\begin{eqnarray}\label{app:ring}
\hat P_{\hexagon} = -{3\over8}+R_2+R_4+R_6, 
\end{eqnarray}
with
\begin{eqnarray*}
R_2(ijklmn)&=& -\frac{2^2}{8} \Big[  (\pmb S_i \cdot \pmb S_j) +  (\pmb S_i \cdot \pmb S_k) + {1\over2}(\pmb S_i \cdot \pmb S_l)\\&&\ \ \ \ \ \ \ + \ {\rm cyclic}(ijklmn)\Big], 
\end{eqnarray*}
\begin{eqnarray*}
R_4(ijklmn)&=& \frac{2^4}{8}   \Big[  (\pmb S_i \cdot \pmb S_j) (\pmb  S_k \cdot \pmb S_l) -  (\pmb S_i \cdot \pmb S_k) (\pmb  S_j \cdot \pmb S_l)  \\
&&\ \ \ \  + {1\over2}(\pmb S_i \cdot \pmb S_j) (\pmb  S_l \cdot \pmb S_m) + {1\over2}(\pmb S_i \cdot \pmb S_k) (\pmb  S_l \cdot \pmb S_n)\ \\
&&\ \ \ \  - (\pmb S_i \cdot \pmb S_l) (\pmb  S_j \cdot \pmb S_n) - {1\over2}(\pmb S_i \cdot \pmb S_l) (\pmb  S_j \cdot \pmb S_m)\ \\
&&\ \ \ \  +  (\pmb S_i \cdot \pmb S_j) (\pmb  S_k \cdot \pmb S_n) + (\pmb S_i \cdot \pmb S_j) (\pmb  S_k \cdot \pmb S_m)\ \\
&&\ \ \ \  + (\pmb S_i \cdot \pmb S_j) (\pmb  S_l \cdot \pmb S_n) + {\rm cyclic}(ijklmn)\Big], \\
R_6(ijklmn) &=& \frac{3\times{2^6}}{8}  \Big[ {1\over3}(\pmb S_i \cdot \pmb S_j) (\pmb  S_k \cdot \pmb S_l) (\pmb S_m \cdot \pmb S_n)  \\
&&\ \ \ \ \ \ \ \ \ \ \  - (\pmb  S_i \cdot \pmb  S_m) (\pmb  S_j \cdot \pmb  S_n) (\pmb  S_k \cdot \pmb  S_l)  \\
&&\ \ \ \ \ \ \ \ \ \ \  + {1\over2} (\pmb  S_i \cdot \pmb  S_n) (\pmb  S_j \cdot \pmb  S_m) (\pmb  S_k \cdot \pmb  S_l) \\
&&\ \ \ \ \ \ \ \ \ \ \  + {1\over2} (\pmb  S_i \cdot \pmb  S_m) (\pmb  S_j \cdot \pmb  S_l) (\pmb  S_k \cdot \pmb  S_n)  \\
&&\ \ \ \ \ \ \ \ \ \ \  - {1\over 6}(\pmb  S_i \cdot \pmb  S_l) (\pmb  S_j \cdot \pmb  S_m) (\pmb  S_k \cdot \pmb  S_n) \\
&&\ \ \ \ \ \ \ \ \ \ \  + \ {\rm cyclic}(ijklmn)\nonumber \Big],
\end{eqnarray*}
where the index $i,j,k,l,m,n \in \hexagon$.
Its graphical form is shown in Fig.\ref{Ring} in the main text.
Specifically speaking, $R_2$, $R_6$, and $R_4$ correspond to the left upper row, right upper row, and bottom row in Fig.\ref{Ring}, respectively.
Note that the index $(i,j,k,l,m,n)$ sorts clockwise on the hexagon.

\section{Effect of the nonbilinear spin interactions}\label{R4}
\begin{figure}[t]
\includegraphics[width=8.4cm]{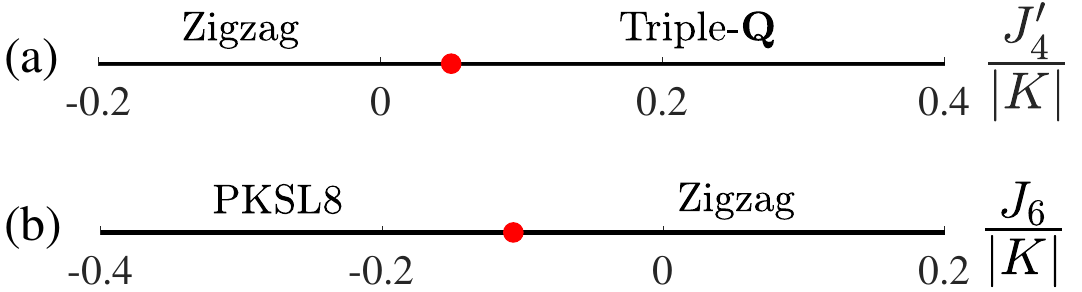}
\caption{(a) Phase diagram of the quantum $K$-$\Gamma$-$\Gamma'$-$J'_{\rm 4}$ model with fixing $\Gamma/|K|=0.4$, $\Gamma'/|K|=-0.05$. 
(b) Phase diagram of the quantum $K$-$\Gamma$-$\Gamma'$-$J_{\rm 6}$ model with fixing $\Gamma/|K|=0.4$, $\Gamma'/|K|=-0.05$. 
}
\label{6spinPD}
\end{figure}

Firstly, among the two-pair interactions, the first and third types are special because they involve two pairs of nearest-neighbor exchanges, as shown in Fig.\ref{Ring}.
We are interested in this special type of interaction because it  
may appear in lower-order perturbations based on the spin-orbital coupled Hubbard model. The specific form of these interactions is given below,

\begin{eqnarray}
H_\text{4-spin}' &&=  J_{\rm 4}'  \sum_{i,j,k,l,m,n \in \hexagon }  \frac{2^4}{8} \Big[  (\pmb S_i \cdot \pmb S_j) (\pmb  S_k \cdot \pmb S_l) \nonumber \\
&&+ {1\over2} (\pmb S_i \cdot \pmb S_j) (\pmb  S_l \cdot \pmb S_m) 
 + \ {\rm cyclic}(ijklmn)  \Big]. 
\end{eqnarray}
A natural question is whether this type of interaction could stabilize the triple-$\bf Q$ order or not. 
Therefore, we consider the following $K$-$\Gamma$-$\Gamma'$-$J'_{\rm 4}$ model containing $K$, $\Gamma$, $\Gamma'$ and $J_{\rm 4}'$ interactions,
\begin{eqnarray}\label{KGGR413}
H'' = H_K+ H_{\Gamma} + H_{\Gamma'} + H_\text{4-spin}'.
\end{eqnarray}
From our VMC calculations of the $K$-$\Gamma$-$\Gamma'$-$J'_{\rm 4}$ model (\ref{KGGR413}) with fixing $\Gamma/|K|=0.4$ and $\Gamma'/|K|=-0.05$, we find that the two-pair spins exchange interaction $J_{\rm 4}'/|K| \gtrsim 0.05$ may stabilize a triple-$\bf Q$ order, as shown in Fig.\ref{6spinPD}(a).
Therefore, these small positive interactions (involving only two Heisenberg interactions between nearest-neighbor sites) certainly stabilize the triple-$\bf Q$ order.

Secondly, we also consider the three-pair spins exchange interaction,
\begin{eqnarray}\label{6spinHami}
&&H_\text{6-spin} = J_{\rm 6} \sum_{i,j,k,l,m,n \in \hexagon } R_6(ijklmn).
\end{eqnarray}
We want to know whether this type of interaction could stabilize the triple-$\bf Q$ order or not. 
From our VMC calculations of the quantum $K$-$\Gamma$-$\Gamma'$-$J_{\rm 6}$ model (i.e., $H'''=H_{\rm K}+ H_{\Gamma} + H_{\Gamma'} + H_\text{6-spin}$) with fixing $\Gamma/|K|=0.4$ and $\Gamma'/|K|=-0.05$, we find that the three-pair spins exchange interaction could not stabilize the triple-$\bf Q$ order, unlike the proposal around the hidden SU(2) point \cite{Lukas2022}. On the one hand, our interaction parameters are far from the hidden-SU(2) point; on the other hand, quantum fluctuations may quantitatively or even qualitatively change the classical order of the ground state.

Interestingly, the negative three-pair spins exchange interaction ($J_{\rm 6}/|K| \lesssim -0.1$) may induce an 8-cone proximate-KSL phase (PKSL8) from a zigzag order, as shown in Fig.\ref{6spinPD}(b).
Note that in a weak magnetic field applied normal to the honeycomb-lattice plane, this PKSL8 state will turn into an Abelian CSL with Chern number $\nu$$=$$2$, which is different from the PKSL8 phase (with $\nu$$=$$-4$ in a small field applied normal to the plane) in Fig.\ref{KGammaGammaPrimeRing} in the main text.

From the above discussions, we find that different kinds of multi-spin exchange interactions may have completely different physical consequences and that the four-spin terms play an important role in stabilizing the triple-$\bf Q$ order. 

In contrast to SU(2) symmetric ring-exchange interactions above, we have also considered the “Kitaev-type” ring-exchange interaction (i.e., Kitaev plaquette operator \cite{Kitaev}), which is anisotropic and breaks the SU(2) rotation symmetry.
However, it turns out that this anisotropic ring-exchange term fails to stabilize the triple-$\bf Q$ order from our VMC calculations.

\section{Construction of trial wave functions}\label{Gutzwiller}
In the VMC framework, the spin interactions in Eq.~(\ref{KGGR}) are rewritten in terms of interacting fermionic operators and are further decoupled into a non-interacting mean-field Hamiltonian $H_{\rm mf}(\pmb R)$, where $\pmb R$ denotes a set of parameters and will be specified in the following.
A series of trial wave functions are constructed by Gutzwiller projection to the mean-field ground state $|\Psi_{\rm mf} (\pmb R)\rangle$.

\subsection{Spin-liquid ansatzes based on PSG}\label{SM:MFs}
A spin-liquid ground state preserves the entire space group symmetry whose point group is $G=D_{3d}\times Z_2^T$. However, the symmetry group of a spin-liquid mean-field Hamiltonian is the projective symmetry group (PSG) \cite{igg,You_PSG} whose group elements are space group operations followed by SU(2) gauge transformations.

It turns out that there are more than 100 classes of PSGs for Z$_2$ QSLs (where the SU(2) gauge symmetry breaks down to the $Z_2$ subgroup in the mean-field Hamiltonian) respecting the $G$ symmetry\cite{You_PSG}. 
We believe that only a few PSGs are relevant to our spin model in all PSGs.
Therefore, in our VMC calculations, we just consider a few of them which are close to the one which describes the symmetry of the exact ground state of the pure Kitaev model.
Here `close' means that the new PSGs and the Kitaev PSG have similar patterns of symmetry fractionalization, namely, they differ by only one or two invariants. The reason for choosing these PSGs is based on the fact that the model (\ref{KGGR}) contains Kitaev interactions. Therefore, it is reasonable to adopt the PSGs that are close to the `Kitaev PSG' given that the non-Kitaev interactions are not extremely large.

Now we provide details of constructing $H_{\rm mf}$ for a given PSG. 
It is convenient  to introduce the matrix operator $\psi_i=( C_i, \bar C_i)$ with $\bar C_i=(c_{i\dn}^\dag, -c_{i\up}^\dag)^T$ such that the spin operators can also be written as $S^m = {\rm Tr}(\psi_i ^\dag {\sigma^m \over4}\psi_i)$. In this form, it is easily seen that there is a local SU(2) gauge symmetry in the fermionic representation of spins\cite{Anderson1988}.
The most general expression of the mean-field Hamiltonian ansatz\cite{Wang2020,PKSL,Liu_KG,Aniso,Li2021}  with nearest neighbor couplings reads,
\beq\label{MFPSG}
H_{\rm mf}^{\rm SL} & = & \!\!\!\! \sum_{\langle i,j \rangle \in\alpha\beta(\gamma)} \!\!\! {\rm Tr}
\, [U_{ji}^{(0)} \psi_i^\dag \psi_j] \! + \! {\rm Tr} \, [U_{ji}^{(1)} \psi_i^\dag
(i R_{\alpha\beta}^\gamma) \psi_j] \nonumber\\ & & \;\;\;\; + {\rm Tr} \, [U_{ji}^{(2)}
\psi_i^\dag \sigma^\gamma \psi_j] \! + \! {\rm Tr} \, [U_{ji}^{(3)} \psi_i^\dag
\sigma^\gamma R_{\alpha\beta}^\gamma \psi_j] \! + \! {\rm H.c.} \nonumber\\
& & \;\;\;\; +  \sum_i {\rm Tr}  (\pmb \lambda_i\cdot  \psi_i \pmb \tau\psi_i^\dag ),
\eeq
where $R_{\alpha\beta}^\gamma = - \frac{i}{\sqrt{2}} (\sigma^\alpha + \sigma^\beta)$,  $\lambda^{x,y,z}$ are three Lagrangian multipliers to ensure SU(2) gauge invariance (where $\lambda^z$ is the one for the particle number constraint), $\tau^{x,y,z}$ are generators of the SU(2) gauge group, and the matrices $U_{ji}^{(0,1,2,3)}$ can be expanded with the identity matrix and $\tau^{1,2,3}$ where the expanding coefficients form a subset of $\pmb R$.  Generally, the values of $\lambda^{x,y,z}$ are zero if there are no external magnetic fields. 

As shown in the following, the PSG constrains the values of the matrices $U_{ji}^{(0,1,2,3)}$.
The gapless KSL is believed to be a finite, stable phase in the presence of non-Kitaev interactions, including the $\Gamma$ and $\Gamma'$ terms.  The mean-field Hamiltonian describing the generic states around the KSL, which we denote the KSL, will then respect the same PSG as the KSL itself. Besides translation symmetry, the symmetry group of the pure KSL, $G = D_{3d}$$\times$$Z_2^T$, has the three generators
\[
S_6 = (C_3)^2 P,\ \ \  M = C_2^{x-y} P,\ \ \   T = i\sigma^y K,
\]
where $C_3$ is a threefold rotation around the direction along $(\pmb x+\pmb y+\pmb z)$, $C_2^{x-y}$ is a twofold rotation around $(\pmb x-\pmb y)$ direction, and $P$ is spatial inversion. 
The PSG of the KSL (called Kitaev PSG) is read most simply from the Majorana
representation, in which the mean-field Hamiltonian is
\begin{eqnarray}\label{Kitaevmf}
H_{\rm mf}^{K} & = & \sum_{\langle i,j \rangle \in\alpha\beta(\gamma)} \rho_a (ic_ic_j) +
\rho_c (i b_i^\gamma b_j^\gamma)  \\ & = & \sum_{\langle i,j \rangle \in\alpha\beta(\gamma)}
i \rho_a {\rm Tr} \left(\psi_i^\dagger \psi_j + \tau^x \psi_i^\dagger
\sigma^x \psi_j + \tau^y \psi_i^\dagger \sigma^y \psi_j \right. \nonumber \\ & &
\;\;\;\;\;\;\; \left. + \tau^z \psi_i^\dagger \sigma^z \psi_j \right) + i \rho_c {\rm Tr}
\left(\psi_i^\dagger \psi_j + \tau^\gamma \psi_i^\dagger \sigma^\gamma \psi_j \right. \nonumber
\\ & & \;\;\;\;\;\; \left. - \tau^\alpha \psi_i^\dagger \sigma^\alpha \psi_j - \tau^\beta
\psi_i^\dagger \sigma^\beta \psi_j\right) + {\rm H.c.} \nonumber
\end{eqnarray}
Because the $c$ fermion never mixes with any of the $b^m$ fermions, any PSG
operation leaves the $c$ fermions invariant. The gauge operation, $W_i(g)$,
following the symmetry operation $g$ should then be $W_i(g) = \pm g$.
A detailed analysis \cite{You_PSG,PKSL,Wang2020} shows that the gauge transformations of the generators $S_6$, $M$, and $T$ are
\beq\label{PSKSL}
W_A(S_6) & =& - W_B(S_6) = \exp \! \big[-i {\textstyle \frac{4}{3}} \pi
{\textstyle \frac{1}{2\sqrt{3}}} (\tau^x + \tau^y + \tau^z) \big],
\nonumber \\
W_A(M) & =& - W_B(M) = \exp \! \big[-i \pi {\textstyle \frac{1}{2\sqrt{2}}}
(\tau^x - \tau^y) \big], \nonumber\\
W_A(T) & = &- W_B(T) = i \tau^y,
\eeq
where A and B denote the two sublattices of the honeycomb lattice.

When the Kitaev honeycomb model is extended to the $K$-$\Gamma$-$\Gamma'$-$J_{\rm R}$ model,
there are several different ansatzes for states beyond the Kitaev
mean-field Hamiltonian [Eq.~(\ref{Kitaevmf})] that are invariant under the
same PSG. The $\Gamma$ interaction gives rise to the mean-field terms
\beq\label{Gm}
H_{\rm mf}^{\Gamma} & = & \!\! \sum_{\langle i,j \rangle \in\alpha\beta(\gamma)} \!\! i\rho_d
(b_i^\alpha b_j^\beta + b_i^\beta b_j^\alpha) \\ & = & \!\! \sum_{\langle i,j \rangle \in\alpha\beta(\gamma)} \!\! i\rho_d {\rm Tr} \left( \tau^\alpha \psi_i^\dag
\sigma^\beta \psi_j + \tau^\beta \psi_i^\dag \sigma^\alpha \psi_j\right)
 + {\rm H.c.} \nonumber
\eeq
and similarly for the $\Gamma'$ interaction
\beq\label{Gmp}
H_{\rm mf}^{\Gamma'} & = & \!\! \sum_{\langle i,j \rangle \in\alpha\beta(\gamma)} \!\! i\rho_f
(b_i^\alpha b_j^\gamma + b_i^\gamma b_j^\alpha + b_i^\beta b_j^\gamma + b_i^\gamma b_j^\beta) \\
& = & \!\! \sum_{\langle i,j \rangle \in\alpha\beta(\gamma)} \!\! i\rho_f {\rm Tr} \left( \tau^\alpha \psi_i^\dag
\sigma^\gamma \psi_j + \tau^\gamma \psi_i^\dag
\sigma^\alpha \psi_j \right. \nonumber \\
&  &\phantom{=\;\;} \left. + \tau^\beta \psi_i^\dag \sigma^\gamma \psi_j + \tau^\gamma \psi_i^\dag \sigma^\beta \psi_j \right) +  {\rm H.c.}  \nonumber
\eeq
For the multi-electron ring-exchange interaction ($J_{\rm R}$),
\beq\label{JR}
H_{\rm mf}^{J_{\rm R}} = \sum_{\langle i,j \rangle }
i \rho_r {\rm Tr} \left(\psi_i^\dagger \psi_j \right) + {\rm H.c.}
\eeq

Comparing with  the general form Eq.~(\ref{MFPSG}),  the decouplings expressed in
Eqs.~(\ref{Kitaevmf}), (\ref{Gm}), (\ref{Gmp}), and (\ref{JR}) contribute the terms
\begin{equation}
\begin{aligned}
& {\tilde U}_{ji}^{(0)} = i (\rho_a + \rho_c + \rho_r), \\
& {\tilde U}_{ji}^{(1)} = i (\rho_a - \rho_c  + \rho_d + 2\rho_f) (\tau^\alpha + \tau^\beta),
\\ & {\tilde U}_{ji}^{(2)} = i (\rho_a + \rho_c) \tau^\gamma + i \rho_f(\tau^\alpha +\tau^\beta ), \\
& {\tilde U}_{ji}^{(3)} = i (\rho_c - \rho_a - \rho_d) (\tau^\alpha - \tau^\beta ),
\end{aligned}
\end{equation}
to the coefficients $U_{ji}^{(m)}$, in which $j$ and $i$ specify $\gamma$.
However, the most general coefficients preserving the $C_3$ rotation symmetry
(in the PSG sense) also contain multiples of the uniform ($I$) and $\tau^x +
\tau^y + \tau^z$ gauge components,
\begin{equation}
\begin{aligned}
& {\tilde {\tilde U}_{ji}^{(0)}} = i \eta_0 + \eta_1 (\tau^x + \tau^y + \tau^z), \\
& {\tilde {\tilde U}_{ji}^{(1)}} = \eta_2 + i \eta_3 (\tau^x + \tau^y + \tau^z), \\
& {\tilde {\tilde U}_{ji}^{(2)}} = \eta_4 + i \eta_5 (\tau^x + \tau^y + \tau^z), \\
& {\tilde {\tilde U}_{ji}^{(3)}} = \eta_6 + i \eta_7 (\tau^x + \tau^y + \tau^z).
\end{aligned}
\end{equation}
If the full symmetry group, $G = D_{3d} \times Z_2^T$, is preserved, then only three parameters $\eta_0$, $\eta_3$, and $\eta_5$ are allowed; by contrast, if one allows the breaking of spatial inversion symmetry, while still preserving mirror reflection symmetry, then $\eta_1$, $\eta_2$, and $\eta_4$ are also allowed. Thus a spin-liquid ansatz that preserves the full PSG symmetry generated by Eq.~(\ref{PSKSL}) contains the variables $U_{ji}^{(m)} = {\tilde U}_{ji}^{(m)} + {\tilde {\tilde U}_{ji}^{(m)}}$ with eight real parameters including $\rho_a$, $\rho_c$, $\rho_d$, $\rho_f$, $\rho_r$, $\eta_0$, $\eta_3$, and $\eta_5$. 
The KSL and PKSL8 both belong to the Kitaev PSG class.
To obtain a reliable phase diagram, we also consider other PSGs which are close to the Kitaev PSG (for more details see Ref.~\onlinecite{Wang2020}).

\subsection{Magnetically ordered states}
To describe the magnetic order of the spin-symmetry-breaking phases of the $K$-$\Gamma$-$\Gamma'$-$J_{\rm R}$ model, we introduce the classical order under single-${\pmb Q}$ approximation\cite{singleQ}
\[\pmb{M}_i = M \{\sin \phi [\hat {\pmb e}_x \cos (\pmb{Q} \cdot
\pmb{r}_i) + \hat{\pmb e}_y \sin(\pmb{Q} \cdot \pmb{r}_i)] + \cos \phi
\, \hat{\pmb e}_z \},\]
where  $\pmb Q$ is the ordering momentum, $\hat {\pmb e}_{x,y,z}$ are the local spin axes (not to be confused with the global spin axes),
and $\phi$ is the canting angle. ($\pi/2-\phi$) describes the angle by which the spins deviate from the plane spanned by $\hat {\pmb e}_x$ and $\hat {\pmb e}_y$. The classical ground state is obtained by minimizing the energy of the trial states.

In our VMC calculations, the static order is treated as a background field coupling to the spins as site-dependent Zeeman field; 
hence the complete mean-field Hamiltonian for the $K$-$\Gamma$-$\Gamma'$ model reads
\begin{equation}\label{Order}
H_{\rm mf}^{\rm total} = H_{\rm mf}^{\rm SL} - {\textstyle \frac{1}{2}} \sum_i
(\pmb {M}_i \cdot C_i^\dagger \pmb \sigma C_i + {\rm H.c.})
\end{equation}

The ordering momentum $\pmb Q$ of $\pmb M_i$ in VMC is adopted from the classical ground state or the classical metal stable states (depending on the energy of the projected state). For a given momentum $\pmb Q$, the local axes $\hat {\pmb e}_{x,y,z}$ are fixed as they are in the classical state. Then $M$ and $\phi$ are treated as variational parameters in the VMC framework. 

In VMC calculations, we have considered not only single-$\bf Q$ ordered states but also multi-$\bf Q$ ordered states, for example, the triple-$\bf Q$ order which is formed by superposing three single-$\bf Q$ order parameters. 
Thus the triple-$\bf Q$ order has a three-fold rotation symmetry.

\begin{table*}[t]
\centering
\begin{tabular}{c|c|c|c|c||c|c|c|c|c}
\hline
\hline
Parent state   & $\frac{\Gamma}{|K|}$&  $\frac{\Gamma'}{|K|}$& $\frac{J_{\rm R}}{|K|}$  & $\, \nu \,$ & $\rho_1$ & $\rho_2$ & $\rho_3$ & ($\rho_4$) & GSD \\
\hline
KSL           &  0.1   & -0.05 &  0.1 & $1$    & 0.9972   & 0.9994  & 1.0034   &          & 3  \\
\hline
PKSL8          &  0.4   &  -0.05 &  0.2 & $-4$   & 0.3067   & 0.3802  & 1.0000   & 2.3131   & 4  \\
\hline
Zigzag         &  0.4   &  -0.05 &  0 & /        & 0.0796   & 0.3414  & 0.4764   & 3.1026   & 4  \\
\hline
Triple-$\bf Q$ &  0.4   &  -0.05 &  0.1 & /      & 0.0045   & 0.0107  & 0.0230   & 3.9618   & 1 \\
\hline
\hline
\end{tabular}
\caption{Eigenvalues of the overlap matrices of the ground states of gapped states on a torus. $\nu$ is the mean-field Chern number of the gapless states which are gapped by a small magnetic field $\pmb B \parallel (\pmb x+\pmb y+\pmb z)$. The system size we adopt is 8$\times$8$\times$2. The data for the ordered states are calculated without applying magnetic fields.
}\label{tab:GSD}
\end{table*}

\section{Ground state degeneracy}\label{GSD}
As all know, the fractional statistics of the quasiparticles in gapped QSLs imply a topology-dependent ground state degeneracy (GSD).
Further, the confinement or deconfinement of the Z$_2$ spin liquid is reflected in the GSD of the Gutzwiller projected state when placed on a torus. If the state is Z$_2$ confined (deconfined), then inserting a global Z$_2$ $\pi$-flux in one of the holes results in the same (a different) state. 
Because this process is equivalent to exchanging the boundary conditions of the mean-field Hamiltonian from periodic to anti-periodic, in two dimensions one may construct the four mean-field ground states $|\psi_{\pm\pm} \rangle$, where the subscripts denote the boundary conditions for the $x$- and $y$-directions. After a Gutzwiller projection of these four states to the physical Hilbert space, the number of linearly independent states is equal to the GSD on a torus.

To make sure that field-induced CSLs are nontrivial, we calculate the density matrix of the projected (VMC) states from the wave-function overlap $\rho_{\alpha\beta} = \langle P_G \psi_\alpha | P_G \psi_\beta \rangle = \rho_{\beta\alpha}^*$, with $\alpha,\beta \in \{++,+-,-+,--\}$.
If $\rho$ has only one significant eigenvalue, with the others vanishing, then the GSD is 1, indicating that the Z$_2$ gauge field is confined. If $\rho$ has more than one near-degenerate nonzero eigenvalue, the GSD is nontrivial and hence the Z$_2$ gauge fluctuations are deconfined. In the deconfined phases, if the Chern number is even then from above the GSD is 4; however, if the Chern number is odd, then the GSD is 3 because the mean-field ground state $|\psi_{++} \rangle$ has odd fermionic parity and vanishes after Gutzwiller projection.
The small-field-induced CSL with $\nu$$=$$1$ or $\nu$$=$$-4$ is certainly deconfined, whose GSD information is shown in Table \ref{tab:GSD}.
For the PKSL14 phase in Fig.\ref{4spinPD}, the small-field-induced CSL with $\nu$$=$$5$ is also deconfined whose GSD is three (not shown).

In addition, we find that the triple-$\bf Q$ ordered state becomes confined after Gutzwiller projection (the GSD on a torus is 1) and belongs to the pure magnetically ordered phase. 
However, as shown in Tab.\ref{tab:GSD}, the zigzag order becomes deconfined after Gutzwiller projection (more than one fairly large eigenvalues in the matrix $\rho$), and it may be an exotic phase with the coexistence of Z$_2$ quantum spin liquid and magnetic order.
By removing the zigzag order while keeping all the other variational parameters intact, we find that the potential coexisting state becomes a proximate-KSL with 20 Majorana cones.
Therefore, the phase transition from the KSL to the potential coexisting state is first-order because of discontinuities in the optimal variational parameters.
It is an interesting question: what is the nature of this potential coexisting phase?
We leave it for future study.

\end{document}